\documentclass[11pt,cmr, a4paper]{article}
\usepackage{moriond,epsfig}

\def\be{\begin{equation}}
\def\ee{\end{equation}}
\def\bea{\begin{eqnarray}}
\def\eea{\end{eqnarray}}

\begin{document}
\vspace*{4cm}
\title{FEEDBACK AND THE INITIAL MASS FUNCTION }

\author{ Joseph Silk}
\address{Astrophysics, Department of Physics, University of Oxford\\ Oxford OX1
3RH UK
}
\maketitle\abstracts{
I describe a turbulence-inspired model for the stellar initial mass function
which includes feedback and self-regulation via protostellar outflows. A new
aspect of the model provides predictions of the star formation rate in
molecular clouds and gas complexes.  A similar approach is discussed for
self-regulation on kiloparsec scales via supernova input, and an expression
is presented  for the global star formation rate that depends on the turbulent
pressure of the interstellar medium.
} %
\noindent
{\small¥{\it Keywords}: star formation-galaxies-turbulence}


\def\simlt{\lower.5ex\hbox{$\; \buildrel < \over \sim \;$}}
\def\simgt{\lower.5ex\hbox{$\; \buildrel > \over \sim \;$}}
\def\simpropto{\lower.2ex\hbox{$\; \buildrel \propto \over \sim \;$}}

\section*{Introduction}
The turbulence paradigm seems to be increasingly accepted in star formation
theory. The dissipation time of the observed
supersonic turbulence especially
in giant molecular clouds (GMCs) is
less than the cloud lifetime, demonstrating that cloud support is by
turbulent pressure. Implications include the consequences that molecular clouds on all
scales are short-lived and
 hence  that large-scale star formation is relatively inefficient.  This
corresponds to what is observed: about 0.1-1
 percent of a GMC is in stars, and the dynamical time-scale  of a GMC is
 $10^6-10^7$ yr,
giving consistency with the global star formation rate of the Milky
Way galaxy, that
requires about 10 percent efficiency of conversion of gas to stars over a
galactic rotation time. On small scales, such as cloud cores, 
supersonic turbulence is still present, although less dominant, and 
one generally  
observes a much higher efficiency of star formation. 
The efficiency  for star clusters embedded in molecular clouds, which
have ages less than $5\times 10^6$ yr, is 10-30 \% (Lada and Lada 2003).
Presumably an important difference is that self-gravity plays a
prominent role on 
these scales.
The turbulence drivers are not well understood. Possible sources
include external drivers (galactic rotation, Parker instability,
OB wind/SN-driven 
super-bubbles) as well as internal sources (protostellar jets and outflows,
gravitational collapse), and the turbulence is likely to be multi-scale,
controlled both by cascades and inverse cascades.

I will focus here mostly on
protostellar outflows on small scales, although I will also discuss
supernova input  on larger scales, as sources of interstellar turbulence. 
The aim will be to demonstrate that
outflow-generated turbulence allows self-regulation of star
formation via control of the accretion rate. A GMC is viewed as a
network of overlapping, interacting  protostellar-driven outflows. This leads to
 simple derivations of the IMF and of the star formation rate.

Similar ideas are implicit in models of competitive  accretion by
turbulent fragments (Bonnell, Vine and Bate  2004). 
The idea of a network of interacting wind-driven shells is 
a manifestation of the fragmentation models
developed by Klessen and collaborators, in which colliding,  supersonic flows  shock,
dissipate and produce dense gravitationally unstable cores (Klessen,
Heitsch and MacLow 2000; Burkert 2003).

\section{Observational issues}

Substantial churning of molecular clouds by protostellar outflows is
a common phenomenon. Observations of some cloud complexes such as Circinus, Orion  and NGC
1333 support  the idea
that protostellar outflows drive the observed turbulence  (Yu, Bally
and Devine 1997; Bally et
al. 1999; Bally and Reipurth 2001; Sandell and Knee 2001).
Quantitatively, such flows seem to be 
an important source
of turbulence. One may estimate the outflow strength as (e.g., Konigl 2003)
$$\dot M_{wind}\sim (r_d/r_A)^2 \dot M_{acc}\sim 0.1   \dot M_{acc} \sim
0.1 v_{amb}^3/G, $$
where the wind velocity is of order 100 km/s, and $v_{amb} \sim 0.3 -3 \rm km\,s^{-1}.$ In general, one expects that 
 $v_{wind}>> v_{amb}.$ Moreover, 
outflow momenta accumulate since typical outflow durations are 
$ t_{wind}\sim 10^5 \rm yr,$ so that the cloud lifetime
$t_{dyn} >> t_{wind} .$ Hence one wind outflow event per forming star
ejecting 10 percent of its mass
can, on the average,  balance the accretion momentum
in a typical cloud core that forms stars at $\sim 10\%$ efficiency.

An important issue is the extent to which the outflows 
are localised. Observed jets suggest that some outflows deposit energy
far 
from the dense cores, but this inference cannot easily be generalised since it
is likely to be highly biased by 
extinction. 
Jets may be ubiquitous, but could still be unstable enough
to drive bipolar outflows and  mostly deposit their 
 energy  in 
and near dense cores. Protostellar outflows are a possible source of 
the observed molecular cloud turbulence, and the only one that is
actually observed 
in situ. The relevant driving scales cover a wide range, and may
well yield the observed, apparently scale-free cloud turbulence. 

The projected density profiles of some dense cores are well  fit by
a  pressure-confined self-gravitating  
isothermal sphere (Alves,  Lada and Lada  2002). 
Others require supercritical cores undergoing subsonic collapse (Harvey et al. 2003).
Magnetic support offers one explanation of the initial conditions,
 with magnetically
 supercritical cores surrounded by subcritical envelopes.
However the core profiles  can also,
more debatably, be reproduced as  supersonic turbulence-compressed
eddies
(Ballesteros-Paredes, Klessen and Vazquez-Semadini 2003).

\section{IMF Theory}

There are many explanations that yield a power-law IMF,
 $mdN/dm=Am^{-x},$ with $x\approx 4/3$, the Salpeter value.  However 
it is harder to account for  the
interesting physics embedded  in $A(p, t...),$ which determines the star formation
rate and efficiency. It is likely that molecular clouds are controlled by
feedback. Protostellar outflows feed the turbulence which controls ambient
pressure, which in turn regulates core formation.  Pressure support includes
magnetic fields.  There are different views about the details of self-regulation,
depending on the relative roles of magnetic and gravitational support versus 
turbulent
compression.  Collapse of
the supercritical cores into protostellar disks   might drive outflows that levitate 
the envelopes, thereby limiting core masses (Shu et al
2003). Ambipolar diffusion and magnetic reconnection would  be
 controlled by the 
turbulence, thereby setting the mass scale of cores 
(Basu \& Ciolek 2004).
The outflows should
 cumulatively set the ambient pressure that in turn controls  both core masses and the accretion rate at which cores can grow
(Silk 1995).

Consider a simple model for a spherically symmetric outflow 
into a uniform medium that drives a shock-compressed shell until 
either the shell encounters another expanding   shell
or until 
pressure balance with the ambient medium is achieved.
At this point, the shell breaks up into blobs confined by ram
or by turbulent pressure,
with the relevent turbulent velocity being that of the shell at break-up.

Now  the shell radius is 
$R=\left(\dot M_{wind}v_w/\rho_a\right)^{1/4}t^{1/2}   \Rightarrow
R\propto v^{-1}, \ \ \ t\propto  v^{-2}.$
 Here $v=dR/dt$ is the shell velocity.
The protostellar outflows generate a network of
interacting shells that form clumps with a velocity distribution: 
$$N(>v)=4\pi R^3 t\dot N_\ast =
\left(\dot M_{wind}v_w/\rho_a\right)^2 v^{-5}\dot N_{\ast}.$$
To convert to a mass function, I  combine this expression for the velocity distribution of clumps
with the relation between mass $m$ and turbulent velocity (identified
with $v$)
for a clump:  $ m=v^3G^{-3/2}(v/B)$ for magnetically 
supercritical cores or  $ m=v^3G^{-3/2}(vp^{-1/2})$ for  
gravitationally supercritical Bonnor-Ebert cores.
Now $B$ scaling suggests $B\simpropto v\Rightarrow \epsilon=3$
(c.f. Shu et al. 2003).
Alternatively, scaling from  Larson's laws
yields 
$\rightarrow \rho \simpropto v^{-2} \Rightarrow \epsilon=4$.
If I instead generalise  the wind-driven,
approximately  momentum-conserving, shell evolution
to $R\propto t^{\delta},$ the IMF can now be written in the form 
$\rightarrow mdN/dm\propto m^{-x} \ \ \  {\rm with}
\ \ \ x=\frac{3\delta + 1}{\epsilon(1-\delta)}$
 where $R\propto
t^{\delta}$ and $m\propto v^{\epsilon}$.
One finally obtains the following values for $x$: $x=5/3$ if $\delta=1/2$; 
$x=4/3$ if $\delta=3/7 \ \ (\epsilon=3)$, and 
$x=5/4$ if $\delta=1/2$; 
$x=4/3$ if $\delta=13/25 \ \ (\epsilon=4)$.
There seems to be little difficulty in obtaining a Salpeter-like IMF.

However in practice, the IMF is not a simple power-law in mass.
It may be described by a combination of three different power-laws:
 $x=-2/3$ over 0.01 to 0.1 M$_\odot$; $x=1/3$ over 0.1 to 1 M$_\odot$;
$x=4/3$ over 1 to 100 M$_\odot $ (Kroupa 2002).
A scale of around  0.3 M$_\odot$ must therefore
be built into the theory.
One clue may come from the fact that the observed IMF 
 is similar to the mass function of dense clumps
in cold clouds, at least on scales above $\sim 0.3 \rm M_\odot$
(Motte et al. 2001).

\section{Feedback}
I now describe an approach that yields the normalisation of the IMF,
and in particular its time-dependence.
The idea is that the network of
interacting shells must self-regulate, in that star formation provides
 both the source of momentum that drives the shells, and is itself controlled by the cumulative pressure that enhances  clump collapse.
I introduce porosity as the parameter that controls self-regulation, via the overlap of outflows. I define porosity as 
$Q=4\pi R_{max}^3 \dot N_\ast t_{max}.$ Now  
for self-regulation, I expect that $Q\sim 1$. One may 
rewrite the IMF as
$$mdN/dm=Q(m_a/m)^x$$
where 
$ m_a=v_a^3G^{-3/2}(v_a/B_a)$ or  $ m_a=v_a^4G^{-3/2}p_a^{-1/2}$.
Now with  $Q\sim 1$, one can expect self-regulation. However in addition,
one must   require self-gravity to
 avoid clump disruption.
This allows the possibility of either negative or positive feedback.

The predicted IMF is  $mdN/dm=Am^{-x}$. The preceding argument yields $A.$
More generally with regard to $x,$ 
if  $R_{wind}\propto t^{\delta}$ and
$m_{clump}\propto v^3$, we obtain $\delta=2/5,3/7,1/2 \rightarrow x= 2/3,4/3,5/3$. The principal new result is the self-regulation ansatz that yields $A$. We infer that $A\propto Q$ where porosity $Q$ can be
written as $f_{\rm low \ density \  phase}=1-e^{-Q}$. 
The IMF slope is in accordance with observations for plausible choices
of parameters. Of course numerical simulations in 3-D are needed to
make a more definitive calculation of the IMF in the context of the
present model.

Nevertheless, there is one encouraging outcome. Turbulent feedback seems to be
significant  for stars of mass $\simgt 0.3 \rm M_\odot .$ This is an
observed fact, and is attributed to detailed  models that
generally invoke   magnetically-driven accretion disk outflows and
jets. Of course even sub-stellar objects display outflows but the 
outflow rates are dynamically unimportant for the parent cloud (e.g., Barrado et al. 2004).
Such a hypothesis could
help  explain  why a feedback explanation of the IMF naturally selects
a  characteristic mass of $\sim 0.3 \rm M_\odot .$ As the stellar mass
increases,
negative feedback  mediates the numbers of more massive clumps and
stars. The
numbers of  increasingly massive stars fall off according to the power-law
derived here.

\section{Summary of IMF results}

There are 3 crucial components to star formation phenomenology.
These are 
the initial stellar mass function or IMF, the star formation
efficiency or SFE, and the star formation rate or SFR. 
The outflow-driven turbulence model predicts these quantities,
provided we can identify the mass of
 a star with 
$m=\mu v^3G^{-3/2}\rho^{-1/2},$ where
$\mu=p^{1/2}/B \ {\rm or} \ 1,$ for either magnetically or
 pressure-supported clouds.
One then finds that, above the feedback scale,  the 
IMF is  $$mdN/dm\propto Q(v_a/v)^5
=f\mu \rho_a Q{m_a}^{2/3}m^{-5/3}.$$
Here $f$ is a constant of order unity.

The SFR is 
 $$\dot n_\ast=\frac{Q }{R_a^3t_a} =Qv_a^{\frac{3\delta +1}{1-\delta}}
\left(\frac{\rho_a}{\dot M_{wind}v_w}\right)^{\frac{1}{1-\delta}}
\propto \frac{Q \rho_a^{2}}{v_a}.$$
In the final expression, I set $\delta=1/2$ and $\dot M_{wind}=0.1v_a^3/G.$
The SFR
$\simpropto \rho_a^{2}. $ 
This means that the SFR accelerates as the cloud evolves and contracts.
The enhanced dissipation from outflows most likely  results in 
the increase of turbulent density in the cloud,
at least until sufficiently massive stars form  whose energetic
 ouflows, winds and eventual explosions  blow the cloud apart.
Evidence for accelerating star formation in many nearby star-forming regions,
based on pre-main-sequence evolutionary tracks,
is presented by Palla and Stahler (2000). This suggests that a
 ministarburst is a common phenomenon.

The SFE is  
$$ \frac{m^\ast_{char}   \dot n_\ast t_{dyn}}{\rho_a}\sim
\frac{m^\ast_{char}   \dot n_\ast}{G^{1/2}\rho_a^{3/2}}\propto
\frac{Q \rho_a^{1/2}}{v_a}.$$
This scaling  gives an SFE
that is $\sim 10-100$ times larger in
cores than in a GMC, more or less as is observed.

There are a number of unresolved issues. The scale-free nature of the 
observed turbulence in molecular clouds is suggestive of a cascade.
Normally these proceed from large to small scales. With internal protostellar
sources, an inverse cascade must be invoked,
such as could arise via injection of turbulence associated with
jet-driven helical  magnetic fields. Alternatively, the wide range of
jet
and outflow scales  suggests that the driving scale may largely be erased.

Efficient  thermal accretion onto low protostellar mass cores 
coupled with protostellar outflows and turbulent fragmentation,
for
which
$M_J^{turb}\sim 
{\cal{M}}^2 M_J^{therm}$ (Padoan and   Nordlund 2002), will help to imprint  the
characteristic stellar mass scale. This suggests  that magnetic
fields, insofar as they regulate and drive outflows,
 are likely to play an important role in setting the
characteristic stellar mass scale. Moreover, regions of enhanced
turbulence, such as would  be associated with star formation induced
by merging galaxies,
could plausibly have a increased feedback scale and hence a top-heavy IMF.

\section{ A theory for kiloparsec-scale outflows}

I now show that a porosity formulation of outflows can also  lead to a
large-scale burst of star formation. Rather than consider molecular cloud
regions, where the physics is more complex, I discuss a more global 
environment where the physics can be simplified but the essential
ingredient of interacting outflows remains.
Consider a larger-scale version of self-regulated feedback.
I model  a cubic kiloparsec of the interstellar medium, which contains
atomic and molecular 
gas clouds and ongoing star formation. I assume that the dominant energy and momentum to the 
 multiphase interstellar medium is via
 supernovae. One expects self-regulation to lead to
a situation in which the porosity $Q\sim 1.$  The porosity is
initially small, but increases as outflows and bubbles develop.
If it is too large, I argue that molecular clouds are disrupted 
and the galaxy blows  much  of the gas
out of the disk, e.g. via fountains into the halo.  
Star formation is quenched until the gas cools and resupplies the
cold gas reservoir in the disk. Whether the gas leaves in a wind is
not clear; this may occur for dwarf galaxy starbursts, but  cannot
happen for Milky-Way type galaxies as long as the supernova rates are
those assumed to apply in the recent past.
 I further speculate that the feedback is initially positive, in a 
normal galaxy. The outflows drive up the pressure of the ambient gas
which enhances the star formation rate by accelerating collapse of
molecular clouds. The feedback eventually is negative in dwarf galaxies,
once a wind develops. In  more massive galaxies, the ensuing starburst
is only limited by the gas supply.

 To develop a simple model, I  make the following ansatz.
The porosity may be defined by 
 $$Q\sim\left( SN \, bubble \,  rate\right)\times
\left(maximum \, bubble \, 4\mbox{-}volume \right)$$
$$\ \ \ \ \ \ \ \ \ \propto \left( star \
formation \ rate\right)\times\left( {turbulent\  pressure}^{-1.4}\right).$$
Expansion of a supernova remnant  is limited by the ambient pressure,
when it  can be described as a radiation pressure-driven snowplow
with $R_a^3t_a\propto  p_{turb}^{1.4}$ (Cioffi et al. 1988).
Hence 
the star formation rate may be  taken to be $\propto Q  p_{turb}^{1.4},$
and by introducing a new parameter $\epsilon$ may also be written
as $\epsilon\times rotation \, rate\times \, gas\, density .$
What is in effect the global star formation efficiency  is now given
by 
$\epsilon\equiv
\left(\frac{\sigma_{gas}}{\sigma_{f}}\right)^{2.7}$, where 
$\sigma_f\approx 20 \rm km s^{-1}\left(E_{SN}/10^{51}\right)^{1.27}
\left(\rm 200M_\odot/m_{SN}\right)$.  
We expect 
positive feedback at
high gas turbulent velocities. High resolution
numerical simulations of a multiphase medium 
demonstrate that a
starburst
is generated, and that the porosity  formalism describes the star
formation rate (Slyz et al. 2004). 
The porosity formulation yields a star formation rate that
gives a remarkably good fit to the
numerical results. The positive feedback arises from the 
implementation of the derived star formation law
with  star formation rate  proportional to  turbulent gas pressure.
Pressure enhancements are mostly due to shocked gas.
It is interesting to note that a  star formation
law which  favours  shock
dissipation   can more readily account
for the spatial extent of star formation as modelled for interacting
galaxies
(Barnes 2004) than can an expression in which the  star
formation rate 
is only a function of gas density (as in the Schmidt-Kennicutt law).

Porosity may therefore regulate star formation, 
on the physical grounds that porosity can be neither too large nor too small.
If it is too small, the rate of massive star formation (and death)
 accelerates until the porosity increases. 
If the porosity is too large the cloud is blown apart via a wind
and loses its gas reservoir. 
To make the concept of a porosity-driven wind more precise, I write
the disk 
outflow rate as the product of the
 star formation rate, the hot gas volume filling
factor, and the cold gas 
mass loading factor. If the hot gas filling factor
$1-e^{-Q}$ ($Q$ is porosity) is of order 50\%, then  this suggests
that the
outflow rate is of order  the star formation rate.
I emphasize that such a result is plausible but only qualitative: it has yet to be numerically 
simulated  in a sufficiently large box.

One infers  that 
the metal-enriched mass ejected in a wind is generically of order the  mass in stars
formed.
This is similar to what is observed for nearby starbursting dwarf galaxies
(e.g.  NGC 1569: Martin, Kobulnicky and Heckman 2002). 
Observations 
suggest that massive galaxies should have had massive winds in the
past,
in order both to account for the observed baryonic mass and the galaxy
luminosity function (Benson et al. 2003), although 
theory has difficulty in rising to this challenge.
It is clear that, energetically,  with conventional supernova rates, one cannot
drive winds from massive or even Milky Way-like  galaxies (Springel and Hernquist
2003). The situation is very different for dwarfs, where supernova input
suffices to drive vigorous winds, although even in these cases
geometric considerations are important (MacLow and Ferrara 1999).

Simulations in a multiphase medium currently lack
sufficient resolution to adequately treat such instabilities as
Rayleigh-Taylor and Kelvin-Helmholtz, that will respectively
enhance the porosity and the wind loading. The highest resolution
simulations to date (Slyz et al. 2004) of a multiphase medium already
show that SN energy input efficiency is considerably underestimated by
failure to have adequate resolution to track the motions of
OB stars from their birth sites in dense clouds before they explode.

It is likely therefore that  feedback may occur considerably
beyond the scales hitherto estimated (Dekel and Silk 1986),
possibly extending to the  galactic   (stellar) mass scale of about
$3\times 10^{10}\rm M_\odot$,  only above which  the star formation efficiency
is inferred to be approximately constant (Kauffmann et al. 2003). 

 Whether even more refined and detailed hydrodynamical simulations can
be consistent with the requirement of substantial early gas loss from
massive galaxies is uncertain (Silk 2003).  One simply lacks the
energy input. Instead, recourse must be made either to an early
top-heavy IMF or  outflows from
a quasar phase that coincided with the epoch of bulge formation,
A top-heavy IMF is motivated by the 
earlier derivation of an IMF driven by turbulent feedback.
The case for an early quasar phase 
during galaxy bulge formation is motivated by the empirical bulge-supermassive
black hole correlation, high quasar metallicities and SMBH growth
times (Dietrich and Hamann 2004).
Yet another option is  an enhanced early rate of hypernovae in starbursts,
as suggested by the interpretation of
 the peculiar abundances found in  the starburst galaxy M82 (Umeda
et al. 2002).
 
While
all of these enhanced sources of energy and momentum are likely to
play some  role in forming galaxies, it is intriguing to note that early
reionisation, in concordance with requirements from CMB measurements
by the WMAP satellite, can also be accomplished by the first of these
hypotheses which can simultaneously account for chemical evolution of
the metal-poor IGM and the abundance ratios observed in extreme
metal-poor halo stars (Daigne et al. 2004).  Moreover, a top-heavy
IMF, if identified with luminous starbursts, can also account for the
faint sub-millimetre galaxy counts (Baugh et al. 2004) and the
chemical abundances in the enriched intracluster medium (Nagashima et
al. 2004).

\def\mnras{MNRAS}
\def\araa{ARAA}
\def\apj{ApJ}
\def\aj{AJ}
\def\pasp{PASP}
\def\apjl{ApJ}

\end{document}